\documentclass[aip, jcp, notitlepage, floatfix, reprint,
  citeautoscript, twocolumn]{revtex4-1}

\usepackage{amsmath}
\usepackage{amsfonts}
\usepackage{graphicx}
\usepackage{dcolumn}
\usepackage[table]{xcolor}
\usepackage[version=3]{mhchem}
\usepackage{transparent}
\usepackage{gensymb}
\usepackage{tabularx}
\usepackage{etoolbox}
\usepackage[hidelinks]{hyperref}
\usepackage{bm}
\usepackage{relsize}
\usepackage[normalem]{ulem}
\usepackage{cases}

\def\maxfloatwidth{%
  \ifdim\columnwidth>246.0pt
  300.0pt  \else
  \columnwidth
  \fi
}

\newcommand{\tbf}[1]{\textbf{#1}}

\newcommand{\mrm}[1]{\mathrm{#1}}
\newcommand{\mbf}[1]{\mathbf{#1}}
\newcommand{\tcr}[1]{\textcolor{black}{#1}}

\newcommand{\me}{\mathrm{e}}

\DeclareMathOperator\erf{erf}
\DeclareMathOperator\erfc{erfc}

\definecolor{bgpeach}{rgb}{1.000,0.925,0.850}
\definecolor{fggray}{rgb}{0.384,0.435,0.471}

\begin{document}


\title{Dielectric response with short-ranged electrostatics}

\author{Stephen J. Cox}
\affiliation{Department of Chemistry, University of Cambridge,
  Lensfield Road, Cambridge CB2 1EW, United Kingdom}
\email{sjc236@cam.ac.uk}

\date{\today}

\begin{abstract}
  The dielectric nature of polar liquids underpins much of their
  ability to act as useful solvents, but its description is
  complicated by the long-ranged nature of dipolar interactions. This
  is particularly pronounced under the periodic boundary conditions
  commonly used in molecular simulations. In this article, the
  dielectric properties of a water model whose intermolecular
  electrostatic interactions are entirely short-ranged are
  investigated. This is done within the framework of local molecular
  field theory (LMFT), which provides a well controlled mean-field
  treatment of long-ranged electrostatics. This short-ranged model
  gives a remarkably good performance on a number of counts, and its
  apparent shortcomings are readily accounted for. These results not
  only lend support to LMFT as an approach for understanding solvation
  behavior, but are relevant to those developing interaction
  potentials based on local descriptions of liquid structure.
\end{abstract}

\maketitle

\section{Introduction}
\label{sec:intro}

Understanding the dielectric nature of polar fluids is one
of the principal aims of liquid state theory, and continues to be the
motivation for both experimental
\cite{fumagalli2018anomalously,chen2016electrolytes,shelton2017water}
and theoretical
\cite{zhang2018note,seyedi2019screening,berthoumieux2019dielectric,pluharova2017size,belloni2018screened,schlaich2016water,loche2020universal,remsing2016role,zhao2020response,cox2018interfacial}
investigations. As can be immediately deduced from the fact that the
free energy of a polar system depends on its shape \cite{frolichBook},
the dipolar interactions that define a polar fluid are long-ranged
(LR). This makes their study both fascinating and complicated. The
importance of understanding the dielectric properties of polar liquids
cannot be overstated, as they play a crucial role in phenomena such as
solvation, self-assembly, and transport through membranes and
nanopores
\cite{leikin1993hydration,bashford2000generalized,roux2004theoretical,honig1995classical}. Consequently,
there are broad implications across the biological, chemical, physical
and materials sciences. Of all polar liquids, water stands out owing
to its ubiquity across disciplines. It is the focus of this article.

In addition to experiments, computer simulation is a widely taken
approach to investigate the behavior of liquids at the molecular
level. The computational cost associated with the microscopic
resolution that simulations provide, however, often limits their
application to system sizes far below that of samples investigated
experimentally. As a result, periodic boundary conditions (PBC) are
often employed to mitigate spuriously high degrees of interfacial
curvature and surface-to-volume ratios. How to appropriately account
for the LR nature of electrostatic interactions, and the implications
this has for dielectric properties, has a long history
\cite{LPS1,LPS2,neumann1983dipole,de1986computer,neumann1986computer,neumann1983calculation1,neumann1983calculation2,neumann1984computer,smith1981electrostatic,ballenegger2014communication}. While
certainly not the only method available to deal with electrostatic
interactions under PBC, Ewald summation techniques are now widely
considered the \emph{de facto} standard \cite{fennell2006ewald}. On
the one hand, it is hard to argue against the success that Ewald
approaches have enjoyed as a computational tool. On the other, they
are not especially intuitive, and risk masking simple physical
interpretations of liquid state behavior.

In this article, the dielectric properties of liquid water whose
intermolecular electrostatic interactions are entirely short-ranged
(SR) will be investigated. Specifically, the framework provided by
\emph{local molecular field theory} (LMFT)
\cite{rodgers2008interplay,rodgers2008local} will be exploited in
order to recast LR electrostatic interactions in a mean-field, yet
well controlled, fashion. Aside from demonstrating how LMFT's
performance can be understood within the existing statistical
mechanical framework for polar liquids, the insight obtained from this
study will aide the development of SR intermolecular potentials, which
is often the case with modern machine-learning approaches
\cite{zhang2018deep,cheng2019ab,zuo2020performance,grisafi2019incorporating,grisafi2018symmetry}. It
will also help us to understand when neglect of LR electrostatics
does, and does not, have severe consequences on simulated
observables. Moreover, it seems likely that strong connections exist
between LMFT ideas and classical density functional theory (see
Refs.~\citenum{archer2013relationship} and~\citenum{remsing2016long}
for differing suggestions), and it is hoped that the results that
follow will help the development of such theoretical approaches. The
results presented here also provide further support to LMFT as a
theoretical approach for understanding the solvation of charged
species \cite{remsing2016long,gao2020short}.

\section*{LMFT and the dielectric constant}

LMFT is a statistical mechanical framework based on the
Yvon-Born-Green hierarchy of equations that relate molecular
correlations with intermolecular forces \cite{rodgers2008local}. The
premise of LMFT is that the intermolecular interactions (or a subset)
can be partitioned into SR and LR portions, and that there exists a
mapping to a `mimic' system. This mimic system comprises
intermolecular interactions arising solely from the SR portion, and a
suitably chosen one-body potential; by construction, the average
structure and higher order correlations of the full system are
captured. Although LMFT can be applied more generally
\cite{weeks2002connecting}, only its application to electrostatic
interactions is considered here. Moreover, as detailed derivations
have been given elsewhere \cite{rodgers2008local}, discussion will be
limited to its most salient features.

Let us begin by noting that the Coulomb potential can be separated
exactly into SR and LR contributions,
\begin{equation}
  \frac{1}{r} = \frac{\erfc(\kappa r)}{r} + \frac{\erf(\kappa r)}{r} \equiv v_0(r) + v_1(r),
\end{equation}
where $\kappa^{-1}$ defines the length scale over which $v_0$
decays. This will be familiar to many as the same procedure taken in
Ewald approaches (see
e.g. Ref.~\citenum{ballenegger2014communication}), where $\kappa$ is
chosen to optimize computational efficiency. In contrast, the success
of LMFT relies on a choice of $\kappa$ such that the mimic system
accurately captures the one-body density and correlations of the full
system. In what follows $\kappa^{-1} = 4.5$\,\AA, which has been
previously demonstrated to be a reasonable choice
\cite{rodgers2008local}. Instead of computing LR electrostatic
interactions explicitly, the effects of $v_1$ are accounted for by a
static restructuring potential,
\begin{equation}
  \label{eqn:VR}
  \mathcal{V}_{\rm R}(\mbf{r}) = \mathcal{V}(\mbf{r}) +
  \int\!\mrm{d}\mbf{r}^\prime\,n_{\rm R}(\mbf{r}^\prime)v_1(|\mbf{r}-\mbf{r}^\prime|),
\end{equation}
where $n_{\rm R}$ is the average charge density in the mimic system,
$\mathcal{V}$ is an external electrostatic potential that would be
applied to the full system, and the integral is understood to be taken
over all space. As $\mathcal{V}_{\rm R}$ is to be chosen such that
$n_{\rm R} = n$, where $n$ is the average charge density of the full
system, Eq.~\ref{eqn:VR} defines a self-consistent relationship
between $\mathcal{V}_{\rm R}$ and $n_{\rm R}$. In a `pure' LMFT
approach, Eq.~\ref{eqn:VR} can be solved either by brute-force or by
exploiting linear-response theory \cite{hu2010efficient}. As the focus
of this article is on understanding dielectric properties within the
LMFT framework, here a more pragmatic approach is instead taken: The
self-consistent cycle is `short-circuited' by using $n$ obtained from
a simulation of the full system as the initial, and only, guess
\cite{rodgers2011efficient}. While Eq.~\ref{eqn:VR} has a simple
mean-field form, it is important to stress that it is not derived from
a mean-field ansatz. It represents a controlled approximation provided
that the mimic system is chosen carefully. It should also be noted
that Eq.~\ref{eqn:VR} has been derived with non-uniform systems in
mind, and that for uniform systems, more sophisticated LMFT approaches
exist
\cite{vollmayr2001using,rodgers2009accurate,rodgers2011efficient}. When
considering uniform systems in this study, however, the
\emph{strong-coupling approximation} (SCA) will be used, in which the
integral in Eq.~\ref{eqn:VR} is simply ignored i.e., $\mathcal{V}_{\rm
  R} = \mathcal{V}$.

The central quantity describing the dielectric behavior of materials
is the static dielectric constant $\epsilon$. A natural question thus
arises: Can we expect SCA to accurately capture $\epsilon$ of the full
system? Following Madden and Kivelson \cite{MaddenKivelson1984sjc}, it
is taken as an empirical fact that $\epsilon$ is an intensive material
property, and therefore does not depend on the shape of the sample
under consideration. This provides the freedom to choose any geometry
for which it is convenient to calculate $\epsilon$, including an
infinite system in which boundaries are not present. In this case, it
is well established that
\begin{equation}
  \label{eqn:MKeps}
  \frac{(2\epsilon+1)(\epsilon-1)}{9y\epsilon}
  = 1 + \frac{4\pi\rho}{3}\int_0^\infty\!\mrm{d}r\,r^2h_\Delta(r),
\end{equation}
with $y=4\pi\beta\rho\mu^2/9$, where $\rho$ is the number density,
$\mu$ is the magnitude of the permanent molecular dipole moment, and
$\beta=1/k_{\rm B}T$. ($T$ is the temperature, and $k_{\rm B}$ is
Boltzmann's constant.) $h_\Delta = 3\langle
h(1,2)\bm{\mu}_1\cdot\bm{\mu}_2/\mu^2\rangle_{\bm{\Omega}_1\bm{\Omega}_2}$
is the projection of the total correlation function $h(1,2)$ on to the
rotational invariant $\bm{\mu}_1\cdot\bm{\mu}_2/\mu^2$, where
$\bm{\mu}_i$ is the dipole vector of molecule $i$, and
$\langle\cdots\rangle_{\bm{\Omega}_1\bm{\Omega}_2}$ denotes an
unweighted average over the orientations of molecules 1 and 2. Crucial
to the current study is that, for an infinite system, $h_\Delta\sim 0$
beyond some microscopic distance $\ell_\epsilon$
\cite{caillol1992asymptotic}. Hence, $\epsilon$ is determined by SR
correlations between dipoles, in line with Kirkwood's original
arguments \cite{kirkwood1939dielectric}. By construction, SCA
accurately describes such SR correlations, from which one can infer
that $\epsilon$ is indeed the same as for the full system.

The above argument skirts around a subtle issue that, as will be
discussed in more detail below, manifests itself as an inconsistency
in the response of the SCA system to uniform fields vs. the
$\mbf{k}\to\mbf{0}$ limit of its external susceptibility, where
$\mbf{k}$ is a reciprocal space wavevector. One might therefore
already anticipate problems where inhomogeneous systems are
concerned. Fortunately, the framework provided by LMFT accounts for
such inconsistencies on average. Moreover, it will also be shown that
the fluctuations deviate from those of the full system in a
predictable manner consistent with dielectric continuum theory (DCT).

\section*{Dielectric response of bulk liquid water}

An infinite system is not a realizable object, even in a computer
simulation that employs PBC. Thus while $\epsilon$ can be `determined'
within the SCA framework from arguments based on an infinite sample,
it still needs to be established how measurable quantities like the
polarization response, or fluctuations at zero field, are affected. In
this section, such properties will be investigated for bulk liquid
water under PBC. This is probably the closest realizable system to the
infinite geometry considered above. Nevertheless, it is important to
bear in mind that there is now an implicit `boundary at infinity'
\cite{LPS1,LPS2,caillol1992asymptotic,de1986computer}.

\begin{figure}[!tb]
  \centering
  \includegraphics[width=8.7cm]{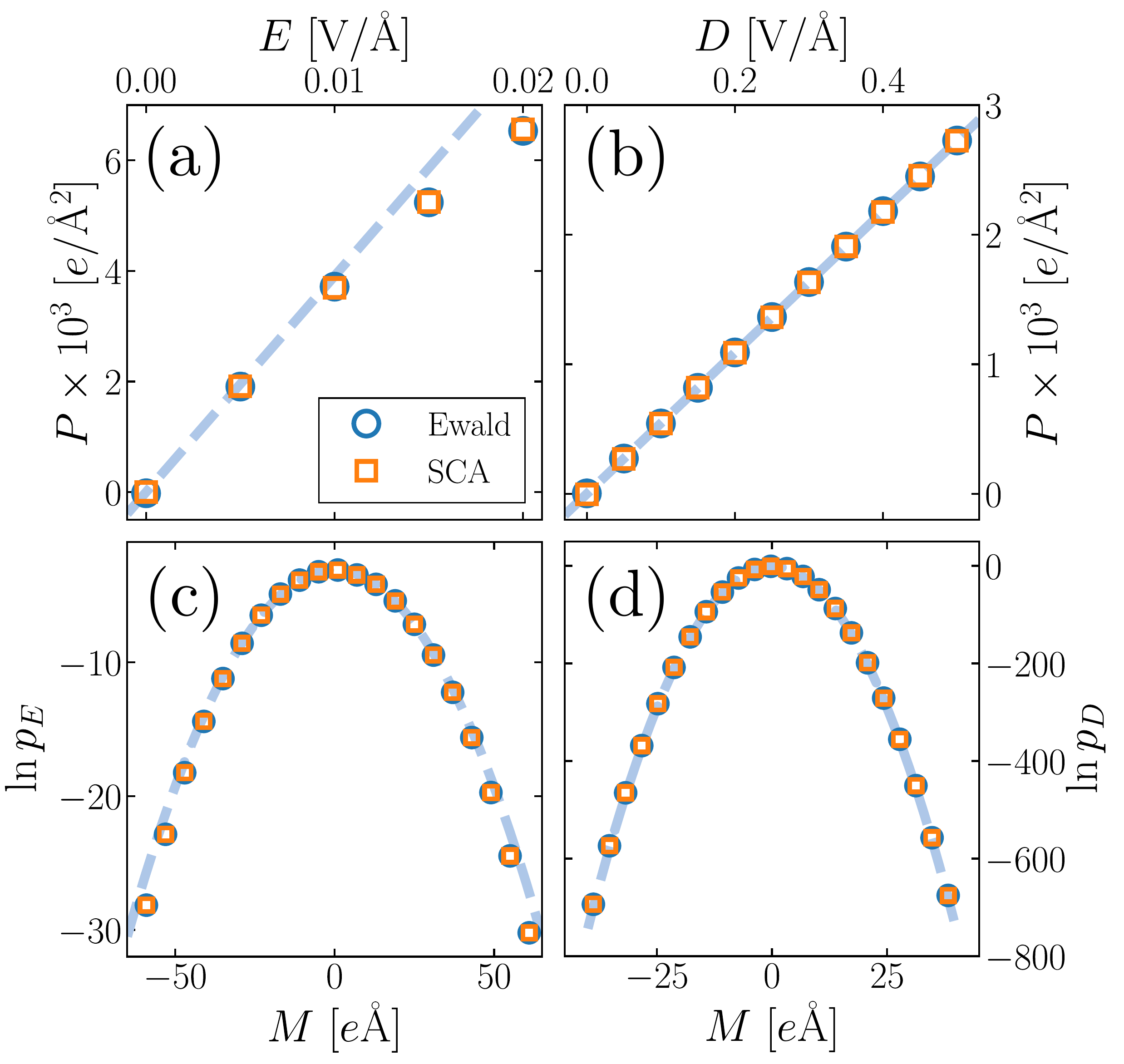}
  \caption{Dielectric response in homogeneous bulk water. (a) and (b)
    show respectively the polarization response $P$ to imposed $E$ and
    $D$ fields (along $z$). (c) and (d) show respectively the
    probability distribution of $M$ at $\mbf{E}=\mbf{0}$ and
    $\mbf{D}=\mbf{0}$. Agreement between SCA and Ewald is
    excellent. Dashed lines indicate the expected response [(a) and
      (b)] or variance [(c) and (d)] from DCT using \tcr{$\epsilon =
      71.7$} (see Fig.~\ref{fig:kirkwood} and
    Eq.~\ref{eqn:CaillolGK}).}
  \label{fig:BulkResponse}
\end{figure}

Figures~\ref{fig:BulkResponse}\,(a) and~(b) show how the polarization
$P$ responds to either a uniform electric field $E$ or electric
displacement field $D$ applied along $z$, respectively (see
e.g. Refs.~\citenum{zhang2016computing1,zhang2016computing2,cox2019finite}). Results
are shown both for the case that LR electrostatics are calculated
explicitly (`Ewald') or neglected entirely (`SCA'). In fact, using SCA
instead of a full LMFT treatment can perhaps be justified here:
$\mathcal{V}_{\rm R} = \mathcal{V}$ on account of the fact that
uniform fields only induce charge density at physical boundaries,
which, at least explicitly, are absent in the current
geometry. ($-\partial_z\mathcal{V} = E$ or $D$ accordingly.) At
constant $E$ a degree of non-linear response is observed at larger
fields, while the response to constant $D$ is linear to an excellent
approximation. In either case, the Ewald and SCA approaches are
virtually indistinguishable over the range of field strengths
studied. This observation is corroborated by the probability
distributions at zero field, $p_E$ and $p_D$, of the $z$-component of
the total dipole moment of the simulation cell $M = \Omega P$, shown
in Figs.~\ref{fig:BulkResponse}\,(c) and~(d). ($\Omega$ is the volume
of the simulation cell.) These have been obtained with histogram
reweighting, and in the case of $p_D$, the wings of the distribution
extend far beyond values of $P$ suggested by
Fig.~\ref{fig:BulkResponse}\,(b). The agreement between the SCA and
Ewald results is remarkable.

The observation that $\langle M^2\rangle$ obtained from SCA agrees
well with Ewald echoes previous studies using Wolf-based
electrostatics \cite{wolf1999exact} to compute $\epsilon$
\cite{armstrong2013water,zahn2002enhancement,yonezawa2012long,fanourgakis2015extension,yonezawa2013electrostatic}. These
studies appealed to the seminal works of Neumann and Steinhauser
\cite{neumann1983dipole,neumann1983calculation1,neumann1983calculation2},
arguing that the fluctuation formula relating $\langle M^2\rangle$ to
$\epsilon$ is largely unaffected when using Wolf-based
approaches. Instead of following a similar strategy here, the
underlying premise of LMFT---to devise a mimic system that accurately
captures the SR correlations of the full system---lends itself more
naturally to analysis in terms of Caillol's results
\cite{caillol1992asymptotic}, which prescribe the asymptotic forms of
the pair correlation functions of polar fluids under different
boundary conditions.

For a cubic simulation cell ($\Omega=L^3$) under PBC, the
electrostatic interactions that enter the Hamiltonian are replaced
with the Ewald potential,
\begin{align}
  \label{eqn:psi-full}
  \psi^{(\lambda)}(\mbf{r}) =
  &\sum_{\mbf{n}}\frac{\erfc(\kappa|\mbf{r}+\mbf{n}L|)}{|\mbf{r}+\mbf{n}L|} \nonumber \\
  + &\frac{1}{L^3}\sum_{\mbf{k}\neq 0}\frac{4\pi}{k^2}\me^{i\mbf{k}\cdot\mbf{r}}\me^{-k^2/4\kappa^2} -
  \frac{2\pi\lambda}{3}\frac{|\mbf{r}|^2}{L^3},
\end{align}
where $\mbf{n}$ is a vector of integers, and $\lambda =
3/(2\epsilon^\prime+1)$. The dielectric constant $\epsilon^\prime$
describes the surrounding medium `at infinity'. In SCA, $\kappa^{-1}$
is chosen sufficiently large such that the sum in reciprocal space can
be ignored without affecting the short-range correlations in the
system,
\begin{equation}
  \label{eqn:psi-lmf}
  \psi^{(\lambda)}_{\rm SCA}(\mbf{r}) =
  \sum_\mbf{n}\frac{\erfc(\kappa|\mbf{r}+\mbf{n}L|)}{|\mbf{r}+\mbf{n}L|}
  - \frac{2\pi\lambda}{3}\frac{|\mbf{r}|^2}{L^3}.
\end{equation}
As Eq.~\ref{eqn:psi-full} is the Green's function for the full system,
it is clear that the charge distribution has been modified by
SCA. This is most easily seen under tin-foil boundary conditions,
$\lambda=0$. In this case $\psi^{(0)}$ describes a periodic set of
unit point charges, each embedded in its own homogeneous compensating
charge, while $\psi^{(0)}_{\rm SCA}$ is instead the Green's function
for a set of periodic unit point charges each embedded in its own
Gaussian compensating charge \tcr{(see Supporting
  Information)}. Crucially, SCA does not alter the boundary condition
at infinity, i.e. the $\lambda$-containing terms in
Eqs.~\ref{eqn:psi-full} and~\ref{eqn:psi-lmf} are identical. Moreover,
the boundary at infinity does not induce structural perturbations,
i.e. $n_{\rm R} = 0$.

The Kirkwood $G$-factor, $G_{\rm K}(r) =
\langle\bm{\mu}_1\cdot\mbf{M}_v\rangle/\mu^2$, describes orientational
correlations between dipoles in the system, where $\bm{\mu}_1$ is a
dipole at the origin, and $\mbf{M}_v$ is the total dipole moment of a
volume $v$. In order to understand SCA's performance under PBC, three
key results from Ref.~\citenum{caillol1992asymptotic} are
required. The first relates $G_{\rm K}$ to $\epsilon$ and
$\epsilon^\prime$,
\begin{equation}
  \label{eqn:CaillolGK}
  yG_{\rm K}(r) = \frac{(2\epsilon+1)(\epsilon-1)}{9\epsilon} +
  \frac{(\epsilon-1)^2}{9\epsilon}\frac{2(\epsilon^\prime-\epsilon)}{2\epsilon^\prime+\epsilon}\frac{v(r)}{L^3}.
\end{equation}
The volume $v$ can either be a sphere of radius $r$, or a cube of
dimension $r$. Setting $v = L^3$ gives the appropriate fluctuation
formula relating $\langle|\mbf{M}|^2\rangle$ and $\epsilon$ for a
given $\epsilon^\prime$. The second is the relation between $G_{\rm
  K}$ and $h_\Delta$,
\begin{equation}
  G_{\rm K}(r) = 1 + \frac{\rho}{3}\int_0^r\!\mrm{d}r^\prime\,\frac{\mrm{d}v}{\mrm{d}r^\prime}h_\Delta(r^\prime).
\end{equation}
The third gives the asymptotic behavior of $h_\Delta$ which, unlike
the infinite system, is now finite,
\begin{equation}
  h_\Delta(r) \sim \frac{1}{3y\rho}\frac{(\epsilon-1)^2}{\epsilon}
  \frac{2(\epsilon^\prime-\epsilon)}{\epsilon(2\epsilon^\prime+\epsilon)}\frac{1}{L^3}.
\end{equation}
This finite asymptotic value is clearly a finite size effect,
vanishing in the limit $L\to\infty$.\footnote{The asymptotic behavior
  of $h_\Delta$ can also be made to vanish by setting
  $\epsilon^\prime=\epsilon$, which effectively samples the infinite
  geometry.} As it has already been argued that both $\epsilon$ and
$\epsilon^\prime$ are unchanged by SCA, it directly follows that
$G_{\rm K}$, and thus $\langle|\mbf{M}|^2\rangle$, are also
unaffected.

\begin{figure}[!tb]
  \centering
  \includegraphics[width=8.7cm]{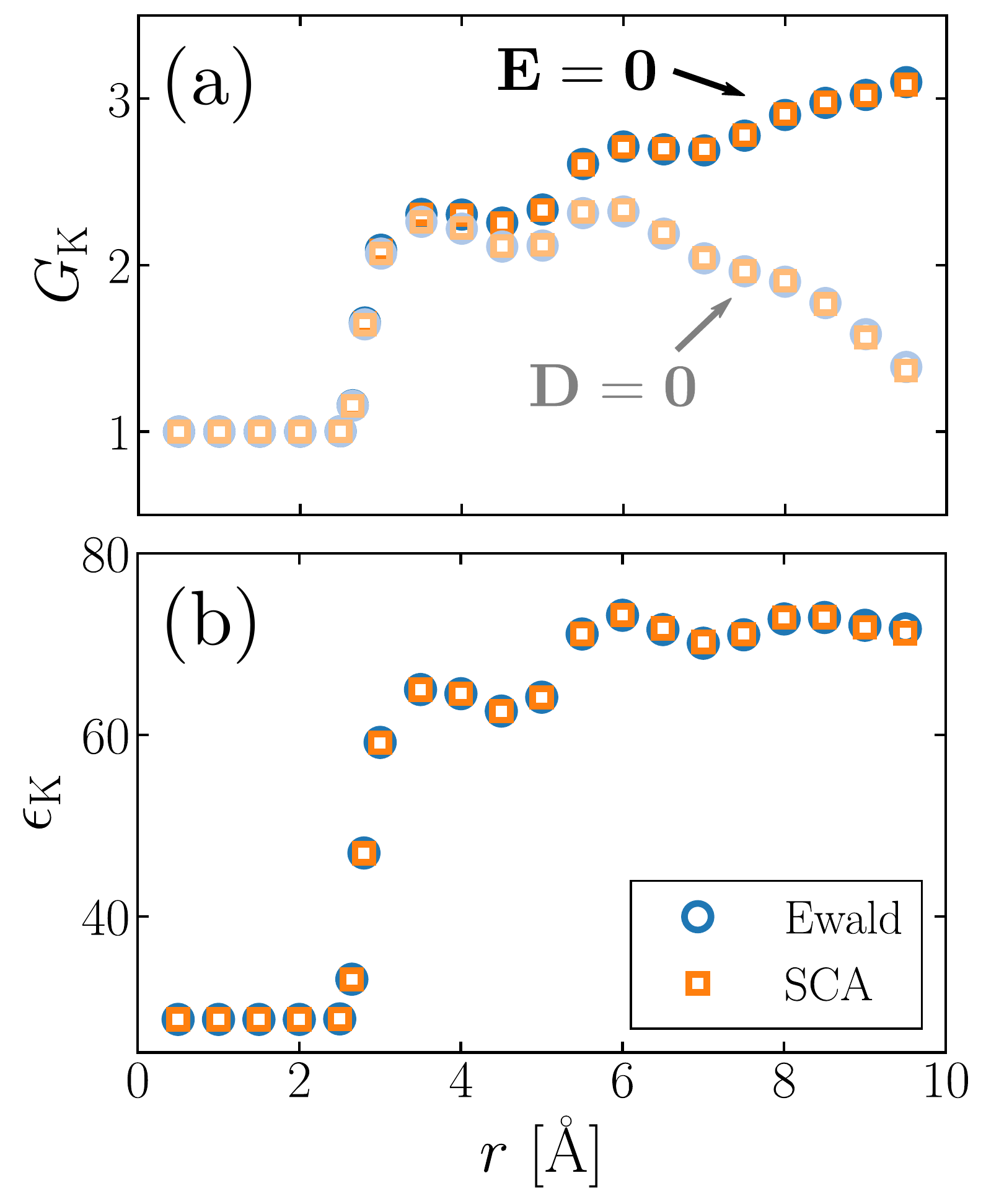}
  \caption{SR orientational correlations determine $\epsilon$. (a)
    Kirkwood $G$-factors obtained at $\mbf{E}=\mbf{0}$
    ($\epsilon^\prime=\infty$) and $\mbf{D}=\mbf{0}$
    ($\epsilon^\prime=0$) have an asymptotic form determined by
    Eq.~\ref{eqn:CaillolGK}. (b) Distance dependent dielectric
    constant $\epsilon_{\rm K}$ obtained from the composite $G$-factor
    (see text). Averaging results for $r>\ell_\epsilon=6$\,\AA{} gives
    \tcr{$\epsilon=71.7$} for both Ewald and SCA.}
  \label{fig:kirkwood}
\end{figure}

The above arguments suggest that, despite $h_\Delta$'s non-vanishing
asymptotic behavior, it can still be considered a SR correlation
function amenable to SCA. Further empirical support for such a notion
is provided by Fig.~\ref{fig:kirkwood}\,(a), where $G_{\rm K}$ is
presented for different spherical subvolumes $v=4\pi r^3/3$. Results
are shown for both $\mbf{E}=\mbf{0}$ and $\mbf{D}=\mbf{0}$,
corresponding to $\epsilon^\prime=\infty$ and $\epsilon^\prime=0$,
respectively \cite{zhang2016computing1,zhang2016computing2}. $G_{\rm
  K}$ obtained from SCA is virtually indistinguishable compared to the
Ewald result. Following Ref.~\citenum{zhang2016computing2}, the
distance dependent dielectric constant $\epsilon_{\rm K}(r)$ can be
found from the asymptotic value of the composite Kirkwood $G$-factor,
$G_{\rm Kc} = (2G_{{\rm K},\mbf{E}=\mbf{0}}+G_{{\rm
    K},\mbf{D}=\mbf{0}})/3$. This result is presented in
Fig.~\ref{fig:kirkwood}\,(b). Averaging $\epsilon_{\rm K}$ for
$r>6\,\text{\AA}=\ell_\epsilon$ gives the macroscopic static
dielectric constant: \tcr{$\epsilon=71.7$} for both the SCA and Ewald
systems. This is in good agreement with existing literature values for
SPC/E water
\cite{van1998systematic,aragones2010dielectric,braun2014transport,zhang2016computing1,zhang2016computing2,cox2019finite}. The
dashed lines in Fig.~\ref{fig:BulkResponse} indicate the expected
response [(a) and (b)] or variance [(c) and (d)] from DCT using
\tcr{$\epsilon = 71.7$.}

\begin{figure}[!b]
  \centering
  \includegraphics[width=8.7cm]{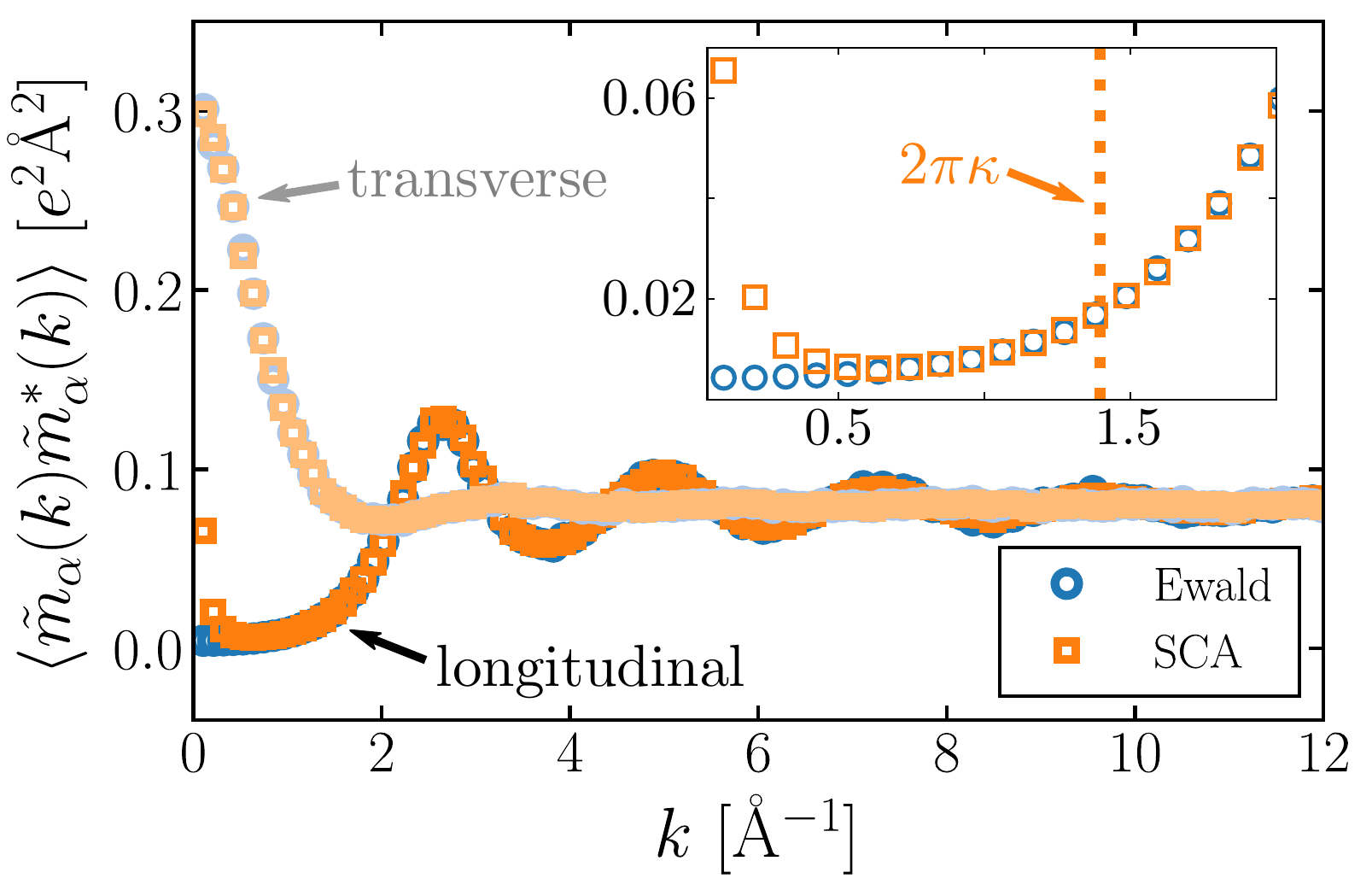}
  \caption{Dipole density correlations in reciprocal space
    $\langle\tilde{m}_\alpha(k)\tilde{m}^\ast_\alpha(k)\rangle$
    determined for both Ewald and SCA electrostatics ($\alpha = x$,
    $y$, $z$, and $\mbf{k}=k\hat{\mbf{z}}$). On the whole, good
    agreement between SCA and Ewald is seen. Inset: At low $k$, SCA's
    longitudinal correlations ($\alpha=z$) deviate from Ewald, tending
    toward the transverse correlations ($\alpha=x$, $y$). These
    deviations occur on a length scale greater than the range
    separation prescribed by SCA, as indicated by the vertical dotted
    line at $k=2\pi\kappa$.}
  \label{fig:mkmk}
\end{figure}

The results presented so far suggest a near flawless performance of
SCA in describing the dielectric properties of polar liquids like
water. While this is entirely consistent with the principle that
$\epsilon$ is determined by SR orientational correlations, it is still
nonetheless remarkable given the history associated with proper
account of LR electrostatics
\cite{LPS1,LPS2,neumann1983dipole,de1986computer,neumann1986computer,neumann1983calculation1,neumann1983calculation2,neumann1984computer,smith1981electrostatic,ballenegger2014communication}. In
fact, it is not immediately obvious that Eqs.~\ref{eqn:MKeps}
and~\ref{eqn:CaillolGK} should hold within the SCA framework: The
factor $(2\epsilon+1)(\epsilon-1)/9\epsilon$ originates from the trace
of the anisotropic external susceptibility
$\bm{\chi}^{(0)}(\mbf{k})\propto\langle\tilde{\mbf{m}}(\mbf{k})\tilde{\mbf{m}}^\ast(\mbf{k})\rangle$
in the $\mbf{k}\to 0$ limit \cite{MaddenKivelson1984sjc}, while for a
system comprising exclusively SR interactions, one would expect
$\bm{\chi}^{(0)}$ to be isotropic at long
wavelengths. ($\tilde{\mbf{m}}$ is the Fourier transform of the
molecular dipole density, using the water oxygen atom as the molecular
center.) Such behavior is indeed hinted at by Fig.~\ref{fig:mkmk},
where $\langle\tilde{m}_\alpha(k)\tilde{m}_\alpha^\ast(k)\rangle$ is
shown, with $\alpha=x,y,z$ and $\mbf{k}=k\hat{\mbf{z}}$. While for the
best part good agreement between the SCA and Ewald approaches is seen,
discrepancies are observed at long wavelengths in the longitudinal
($\alpha=z$) fluctuations, with the SCA results sharply increasing as
$k\to 0$. It is interesting that these deviations appear at length
scales far larger than the range separation prescribed by SCA (see
inset).

In a full treatment of electrostatics, setting $\mbf{E}=\mbf{0}$
ensures $\chi^{(0)}_{xx}$ and $\chi^{(0)}_{yy}$ are continuous at
$k=0$, e.g. $\chi^{(0)}_{xx}(k\to 0) = \chi^{(0)}_{xx}(0)$. On the
other hand, $\chi^{(0)}_{zz}$ is discontinuous
i.e. $\chi^{(0)}_{zz}(k\to 0)\neq \chi^{(0)}_{zz}(0)$
\cite{neumann1986computer}. The situation is reversed for
$\mbf{D}=\mbf{0}$. In contrast, Fig.~\ref{fig:mkmk} suggests that with
SCA, $\chi^{(0)}_{\alpha\alpha}(k\to 0) =
\chi^{(0)}_{\alpha\alpha}(0)$ at $\mbf{E}=\mbf{0}$ and
$\chi^{(0)}_{\alpha\alpha}(k\to 0) \neq \chi^{(0)}_{\alpha\alpha}(0)$
at $\mbf{D}=\mbf{0}$, irrespective of whether $\alpha = x$, $y$ or
$z$. The fact that $\mbf{k}=\mbf{0}$ response to both $\mbf{E}$ and
$\mbf{D}$ in SCA well describes that of the full system therefore
suggests an inconsistency within the SCA framework. We will see the
consequences of this when considering an inhomogeneous system
below. There it will also be shown that the $k\to 0$ longitudinal and
transverse external susceptibilities are indeed equal in SCA i.e. the
former is too large by a factor $\epsilon$. (See also \tcr{Supporting
  Information}.) Fortunately, LMFT readily provides a route to account
for this inconsistency.

\section*{Dielectric response with extended interfaces}

Placing systems under PBC is a useful construction for investigating
bulk properties of materials such as $\epsilon$, especially when
computational resources are limited. Real systems, however, have
boundaries that will induce structural inhomogeneities. These
structural inhomogeneities may also be accompanied by regions of
non-vanishing charge density of the polar liquid. This could arise
from the boundary itself preferentially orienting the molecules of the
liquid (e.g. due to functional groups at a solid surface), or from an
asymmetric charge distribution in the liquid's constituent molecules
(e.g. water). In any event, it is simply not enough to evaluate the
performance of SR interaction potentials on their ability to reproduce
properties of homogeneous systems. Rather, it is imperative to assess
and understand their behavior in the presence of extended interfaces.

\begin{figure}[!tb]
  \centering
  \includegraphics[width=8.7cm]{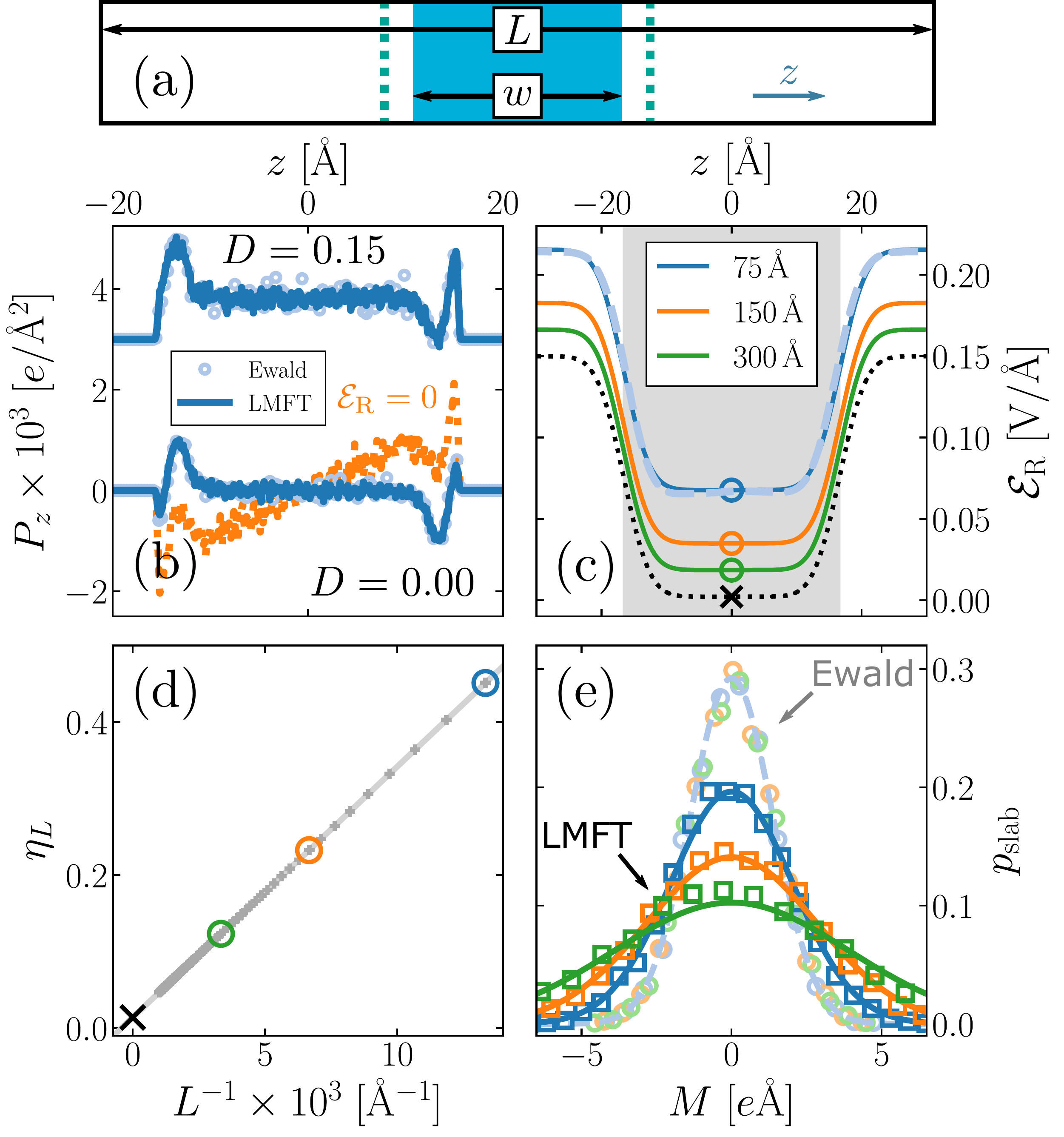}
  \caption{Dielectric response of water confined between hydrophobic
    walls, shown schematically in (a). A water slab of thickness $w$,
    centered at $z=0$, forms two interfaces with vacuum in the
    $xy$-plane at $z=\pm w/2$. $L$ denotes the total length of the
    simulation cell along $z$. The dotted lines depict the confining
    walls. (b) Average polarization profiles $P_z$ at both $D=0$ and
    $D=0.15$\,V/\AA{} (shifted vertically by $3\times
    10^3$\,$e$/\AA$^2$ for clarity) are well described by
    LMFT. Neglecting $\mathcal{E}_{\rm R}$ at $D=0$ results in poor
    agreement, as indicated by the orange dotted line. (c)
    $\mathcal{E}_{\rm R}$ predicted by DCT with
    $\mathcal{E}=0.15$\,V/\AA, \tcr{$\epsilon=71.7$} and
    \tcr{$w=33.2$\,\AA}. Solid lines have been obtained by numerically
    evaluating Eq.~\ref{eqn:ERL} for finite $L$ (see legend). The
    dashed line is a simulation result for $L=75$\,\AA, with the
    spontaneous contribution subtracted. The dotted line shows the
    $L\to\infty$ result (Eq.~\ref{eqn:ERinf}). The shaded region
    indicates $-w/2\le z\le w/2$. (d) $\eta_L$ vs $L^{-1}$ obtained by
    evaluating Eq.~\ref{eqn:ERL} at $z=0$ [see open circles in
      (c)]. The solid line is a linear fit. The black cross indicates
    $\eta_\infty = 1/\epsilon$. (e) Probability distributions of the
    slab's dipole moment. With Ewald, this is relatively insensitive
    to $L$. With LMFT, DCT predicts the variance increases as
    $\eta_L^{-1}$, as indicated by the solid lines.}
  \label{fig:Slab}
\end{figure}

In this section, the dielectric properties of water confined between
structureless, repulsive walls will be investigated. This is a
prototypical model for understanding nanoconfined water in hydrophobic
environments. In such a geometry, the interface is approximately
planar, and Eq.~\ref{eqn:VR} can be recast as
\begin{equation}
  \label{eqn:VR-1D}
  \mathcal{V}_{\rm R}(z) = \mathcal{V}(z) + \frac{1}{L}\sum_{k\neq 0}\frac{4\pi}{k^2}\tilde{n}_{\rm R}(k)\exp(ikz)\exp(-k^2/4\kappa^2),
\end{equation}
where $\tilde{n}_{\rm R}$ denotes a Fourier component of $n_{\rm R}$,
and $L$ is now the total length of the simulation cell in the
direction perpendicular to the interface (taken to be $z$). The system
is still understood to be replicated in all three dimensions. A
schematic is shown in Fig.~\ref{fig:Slab}\,(a).

Even in the absence of an external field, liquid water has a
non-vanishing charge density close to the interface. Consequently,
$\mathcal{V}_{\rm R}$ is finite, along with a corresponding
restructuring field $\mathcal{E}_{\rm R} = -\partial_z\mathcal{V}_{\rm
  R}$. Neglecting $\mathcal{E}_{\rm R}$ has severe consequences for
the orientational statistics of water in a confined geometry. This was
already discussed by Rodgers and Weeks \cite{rodgers2008interplay} and
their results are recapitulated in a slightly different form in
Fig.~\ref{fig:Slab}\,(b), where the average molecular dipole density
along $z$ is shown. While the average polarization obtained with LMFT
agrees well with the Ewald system, the SCA system on its own
($\mathcal{E}_{\rm R}=0$) yields a qualitatively incorrect
picture. Crucial to what follows is that LMFT also gives the correct
average polarization in the presence of a uniform field, which is also
shown in Fig.~\ref{fig:Slab}\,(b) for $D=-\partial_z\mathcal{V} =
0.15$\,V/\AA.

It is clear that LMFT provides a means to correct for the effects of
neglecting LR electrostatics on the average dielectric response in
inhomogeneous systems. Results from previous studies
\cite{hu2014symmetry,pan2017effect,yi2017connections,pan2019analytic,baker2020local}
suggest the fluctuations will also be affected, and establishing how
they are affected is likely to provide useful physical insight. To set
about tackling this issue, let us consider a continuum model in which
a uniform dielectric slab with thickness $w$ is centered at $z=0$ such
that its boundaries occur at $z_\pm = \pm w/2$. A vacuum region exists
either side of the slab. If the slab has a uniform polarization $P$,
this leads to a charge density at the boundaries, $n(z) =
P\left[\delta(z-w/2)-\delta(z+w/2)\right]$. Recalling that $n_{\rm R}
= n$ at self-consistency, taking the Fourier transform of $n$,
substituting into Eq.~\ref{eqn:VR-1D} and differentiating to find
$\mathcal{E}_{\rm R}$ gives
\begin{equation}
  \label{eqn:ERL}
  \mathcal{E}_{\rm R}(z) = \mathcal{E}(z) - \frac{8\pi P}{L}\sum_{k\neq 0}\frac{\cos(kz)\sin(kw/2)}{k}\exp(-k^2/4\kappa^2).
\end{equation}
In the limit $L\to\infty$, this can be solved analytically,
\begin{align}
  \label{eqn:ERinf}
  \lim_{L\to\infty}\mathcal{E}_{\rm R}(z) &= \nonumber \\
  \mathcal{E}(z) -  &2\pi P\bigg\{\erf\big[(w/2-z)\kappa\big] + \erf\big[(w/2+z)\kappa\big]\bigg\}.
\end{align}
In this case it is instructive to consider the limiting
values of $\kappa$,
\begin{subnumcases}{\mathcal{E}_{\rm R}(z)=}
  \mathcal{E}(z)          & (as $\kappa\to 0$), \label{eqn:ERinf_lim0}\\
  \mathcal{E}(z) -4\pi P  & (as $\kappa\to\infty$). \label{eqn:ERinf_limInf}
\end{subnumcases}
The result for $\kappa\to 0$ simply states that all electrostatic
interactions have been accounted for explicitly in the SCA system. In
the case $\kappa\to\infty$, the result can be interpreted as follows:
Due to the neglect of LR electrostatics, the SCA system omits the
depolarizing field established by the induced surface charge density
at the boundaries, which is then accounted for by the second term
($-4\pi P$) in Eq.~\ref{eqn:ERinf_limInf}. For finite $\kappa$, it is
found empirically that $\mathcal{E}_{\rm R} = \mathcal{E} -4\pi P$ is
an excellent approximation in the slab's interior, provided
$w\gg\kappa^{-1}$. In the general case of finite $L$,
Eq.~\ref{eqn:ERL} can be solved numerically in a straightforward
manner. This is shown in Fig.~\ref{fig:Slab}\,(c) for $L = 75$, $150$,
and $300$\,\AA, along with the analytic result (Eq.~\ref{eqn:ERinf})
for $L\to\infty$, using \tcr{$\epsilon=71.7$} obtained above. Also
shown is $\mathcal{E}_{\rm R}$ for $L=75$\,\AA{} obtained from
simulation, with the spontaneous contribution subtracted \tcr{(see
  Supporting Information)}. The simple dielectric continuum model
presented above captures this result from molecular simulation with
remarkable accuracy.

When subjected to an external field along $z$, the slab responds
according to its longitudinal external susceptibility,
\begin{equation}
  \label{eqn:P_dct_full}
  4\pi P = 4\pi\chi^{(0)}_{zz}\mathcal{E} =
  \bigg(\frac{\epsilon-1}{\epsilon}\bigg)\mathcal{E}.
\end{equation}
By construction, LMFT gives the same $P$ as the full system. The field
it responds to, however, is $\mathcal{E}_{\rm R}$ rather than
$\mathcal{E}$,
\begin{equation}
  \label{eqn:P_dct_lmf}
  4\pi P = 4\pi\chi^{(0)}_{{\rm R},zz}\mathcal{E}_{\rm R} \approx
  4\pi\chi^{(0)}_{{\rm R}, zz}\eta_L\mathcal{E},
\end{equation}
where $\eta_L\le 1$ is an $L$-dependent scalar relating
$\mathcal{E}_{\rm R}$ to $\mathcal{E}$, and the relationship is
approximate as variations close to the boundaries are ignored. By
evaluating Eq.~\ref{eqn:ERL} at $z=0$, $\eta_L$ is found to scale
linearly with $1/L$, as seen in Figs.~\ref{fig:Slab}\,(c) and~(d). In
the limit $L\to\infty$, substituting Eq.~\ref{eqn:ERinf_limInf} gives
$4\pi P = 4\pi\chi^{(0)}_{{\rm R},zz}[\mathcal{E}-4\pi
  P]$. Substituting Eq.~\ref{eqn:P_dct_full} and rearranging yields,
\begin{equation}
  \lim_{L\to\infty} 4\pi\chi^{(0)}_{{\rm R},zz} = \epsilon-1.
\end{equation}
Comparing with Eqs.~\ref{eqn:P_dct_full} and~\ref{eqn:P_dct_lmf}, it
is clear that $\eta_\infty = 1/\epsilon$. Thus it is indeed the case
that the longitudinal external susceptibility is too large by factor
$\epsilon$ as $k\to 0$. \tcr{(See also Supporting Information.)}

This simple dielectric continuum model directly elicits information on
the fluctuations in the LMFT formalism. Specifically, it immediately
follows that $\langle M^2\rangle$ in LMFT is a factor $1/\eta_L$
larger that it is with Ewald. This is confirmed in
Fig.~\ref{fig:Slab}\,(e), where the probability distributions of $M$
for the slab $p_{\rm slab}$ obtained from simulations with different
$L$ are plotted along with Gaussian distributions with variances
predicted by this dielectric continuum model. It is stressed that
these are not fits to the simulation data: \tcr{$\epsilon = 71.7$} has
been determined from the simulations of bulk water, and
\tcr{$w=33.2$}\,\AA{} has been determined from the variance of $p_{\rm
  slab}$ obtained using Ewald sums \tcr{(see Supporting
  Information)}. The fact that this simple DCT model describes the
behavior of the LMFT system so well is further support for the notion
that $\epsilon$ is unchanged from that of the system with full
electrostatics.

\section*{Discussion}

LMFT provides an elegant statistical mechanical framework that is
readily compatible with standard molecular simulation approaches, and
can be applied in cases when it is not always clear, at least \emph{a
  priori}, that conventional mean field treatments will
work. Improving our theoretical descriptions of solvation is one area
where LMFT has enjoyed much recent success
\cite{gao2020short,remsing2016long,gao2018role}. This includes both
hydrophobic and ionic solvation across a range of length scales. As
the solvent's dielectric constant plays a central role in our
understanding of solvation, particularly in the case of charged
solutes, the results presented above lend further weight to LMFT as a
suitable approach for investigating solvation, and should facilitate
its development going forward.

Of particular relevance to this work is the recent study of Gao
\emph{et al}, in which LMFT was used to probe ion correlations in
water \cite{gao2020short}. It was found that treating all
electrostatic interactions with the SR $v_0$ led to potential of mean
forces (PMFs) that disagreed with a full electrostatic treatment. This
was attributed to the inability of this `$v_0$-only' approach to
capture effects of dielectric screening on the PMFs. Building on
results from Remsing \emph{et al.} \cite{remsing2016long}, this
discrepancy was corrected by introducing a renormalized direct ion-ion
interaction, which was shown to have an asymptotic limit ($\sim
1/\epsilon r$) consistent with DCT. A pleasing aspect of the results
presented above is that they show this renormalized interaction is
consistent with the static dielectric constant of the underlying SR
solvent. The same can also be said for corrections to thermodynamic
properties of uniform systems, e.g. the internal energy and pressure,
that depend on $\epsilon$ \cite{rodgers2009accurate}.

The fact that certain dielectric properties of polar liquids can be
captured with SR interaction potentials could be highly advantageous
to those seeking to describe liquids without explicit reference to LR
electrostatic interactions. It is abundantly clear, however, that the
effects of LR electrostatics cannot simply be neglected entirely. This
was already obvious from early LMFT studies on liquid water, even in
the absence of external fields
\cite{rodgers2008interplay,remsing2011deconstructing}. It is also
clear that the polarization fluctuations of inhomogeneous systems will
also be affected, which could have been anticipated from previous
studies. In particular, symmetry-preserving mean-field theory (an
extension of LMFT)
\cite{hu2014symmetry,pan2017effect,yi2017connections,pan2019analytic}
likely provides a means to recover the fluctuations by capturing both
equilibrium and dynamical effects of interfaces with high symmetry
(see also Ref.~\citenum{baker2020local}). In this article, no such
attempt to correct the fluctuations in LMFT has been made. Instead, it
has been demonstrated that dielectric properties of a system with SR
electrostatic interactions are described well by DCT where $\epsilon$
is unchanged from that of a system with full electrostatics.

Earlier in this article, it was taken as given that $\epsilon$ is an
intensive quantity that does not depend upon sample shape. While this
is supported by rigorous theoretical calculations (see
e.g. Ref.~\citenum{chandler1977dielectric}), the fact that dielectric
properties can be understood within the LMFT framework can be viewed
as a demonstration of this result, and is perhaps more open to
intuitive physical interpretation. This may prove useful as we
continue to develop our understanding of dielectrics under confinement
\cite{zhang2018note,fumagalli2018anomalously,schlaich2016water,loche2020universal}.

\section*{Methods}
{\footnotesize All simulations used the SPC/E water model
  \cite{BerendsenStraatsma1987sjc}, whose geometry was constrained
  using the RATTLE algorithm \cite{andersen1983rattle}. Dynamics were
  propagated using the velocity Verlet algorithm with a time step of
  2\,fs. The temperature was maintained at 298\,K with a
  Nos\`{e}-Hoover chain
  \cite{shinoda2004rapid,tuckerman2006liouville}, with a damping
  constant 0.2\,ps. Where applicable, the particle-particle
  particle-mesh Ewald method was used to account for long-ranged
  interactions \cite{HockneyEastwood1988sjc}, with parameters chosen
  such that the root mean square error in the forces were a factor
  $10^{5}$ smaller than the force between two unit charges separated
  by a distance of 1.0\,\AA{} \cite{kolafa1992cutoff}. A cutoff of
  10\,\AA{} was used for non-electrostatic interactions: For
  simulations using LMFT/SCA, this cutoff was used for all
  interactions. The LAMMPS simulation package was used throughout
  \cite{plimpton1995sjc}. For simulations with an imposed electric
  displacement field, the implementation given in
  Ref.~\citenum{cox2019finite} was used. Simulations using LMFT/SCA
  required further modification of the LAMMPS source code, \tcr{which
    has been made freely available.}

For results presented in Fig.~\ref{fig:BulkResponse}, the system
comprised 256 molecules in a cubic simulation cell of dimension
$L=19.7304$\,\AA. For Figs.~\ref{fig:BulkResponse}\,(b) and~(d), an
electric displacement field was imposed in all three dimensions
i.e. $\mbf{D}=D_x\hat{\mbf{x}} + D_y\hat{\mbf{y}} + D\hat{\mbf{z}}$
with $D_x=D_y=0$. See Ref.~\citenum{zhang2016computing1} for further
discussion of this point. Simulations at constant $\mbf{E}$ were run
for 150\,ns post equilibration, while those at constant $\mbf{D}$ were
run between 2.5\,ns and 5.0\,ns. The probability distributions $p_E$
and $p_D$ were obtained using the multistate Bennett acceptance ratio
method \cite{ShirtsChodera2008sjc}, with simulations performed at
$E=0,\pm 0.005,\ldots,\pm 0.020$\,V/\AA, and $D = 0,\pm
0.05,\ldots,\pm 1.45$\,V/\AA, respectively. The Kirkwood $G$-factors
(Fig.~\ref{fig:kirkwood}) were obtained from the same set of
simulations, although those at $\mbf{E}=\mbf{0}$ were performed for a
further 150\,ns with configurations stored more frequently (every
30\,ps). Results presented in Fig.~\ref{fig:mkmk} were obtained from
45-50\,ns simulations of 6912 molecules with $L=59.1912$\,\AA{} and
$\mbf{D}=\mbf{0}$.

For simulations of water between hydrophobic walls
(Fig.~\ref{fig:Slab}), 400 water molecules were confined between
Lennard-Jones 9-3 walls,
\begin{align}
  u_{\rm wall}(z) &= \varepsilon_{\rm w}\left[
    \frac{2}{15}\left(\frac{\sigma_{\rm w}}{\Delta_{\rm lo}z}\right)^9
    - \left(\frac{\sigma_{\rm w}}{\Delta_{\rm lo}z}\right)^3 \right] \nonumber \\[5pt]
  &+ 
  \varepsilon_{\rm w}\left[
    \frac{2}{15}\left(\frac{\sigma_{\rm w}}{\Delta_{\rm hi}z}\right)^9
    - \left(\frac{\sigma_{\rm w}}{\Delta_{\rm hi}z}\right)^3 \right],
\end{align}
where $\Delta_{\rm lo}z = |z-z_{\rm lo}|$ and $\Delta_{\rm hi}z =
|z-z_{\rm hi}|$ with $z_{\rm lo} \le z \le z_{\rm hi}$ located within
the primary simulation cell. All simulations used $z_{\rm hi} =
-z_{\rm lo} = 17.25$\,\AA, $\varepsilon_{\rm w} = 0.6$\,kcal/mol and
$\sigma_{\rm w} = 2.5$\,\AA. The potential was truncated and shifted
for $\Delta_{\rm lo(hi)}z\ge 2.1459$\,\AA. Simulations were between
25\,ns and $\sim$90\,ns. Hybrid boundary conditions were used
\cite{zhang2016computing1} i.e. $\mbf{D}=D\hat{\mbf{z}}$ and
$E_x=E_y=0$. For $D=0$, this is formally equivalent to the
Yeh-Berkowitz correction for the slab geometry.}
\cite{YehBerkowitz1999sjc}

\acknowledgments{Michiel Sprik and Rob Jack are thanked for insightful
  discussions. Sim\'{o}n Ram\'{i}rez-Hinestrosa is thanked for
  technical discussions regarding the LAMMPS
  implementation. Computational support from the UK Materials and
  Molecular Modelling Hub, which is partially funded by EPSRC
  (EP/P020194), for which access was obtained via the UKCP consortium
  and funded by EPSRC grant ref EP/P022561/1, is gratefully
  acknowledged. I am supported by a Royal Commission for the
  Exhibition of 1851 Research Fellowship.}

\section*{Data Availability}

Source code beyond the standard LAMMPS distribution used to perform
simulations described here can be accessed at
\url{https://github.com/uccasco/LMFT}. Input files for the simulations
are openly available at the University of Cambridge Data Repository,
\url{https://doi.org/10.17863/CAM.52565}.

\vspace{0.2cm}
\noindent The published version of this article can be found at
\url{https://doi.org/10.1073/pnas.2005847117}.

\bibliography{./pnas/cox.bib}

\clearpage
\onecolumngrid
\renewcommand\thefigure{S\arabic{figure}}
\renewcommand\theequation{S\arabic{equation}}
\setcounter{figure}{0}
\setcounter{equation}{0}

\noindent {\Large \tbf{Supporting Information}}

\section*{Comparing the DCT model with molecular simulation}

\begin{figure*}
  \centering
  \includegraphics[width=11.4cm]{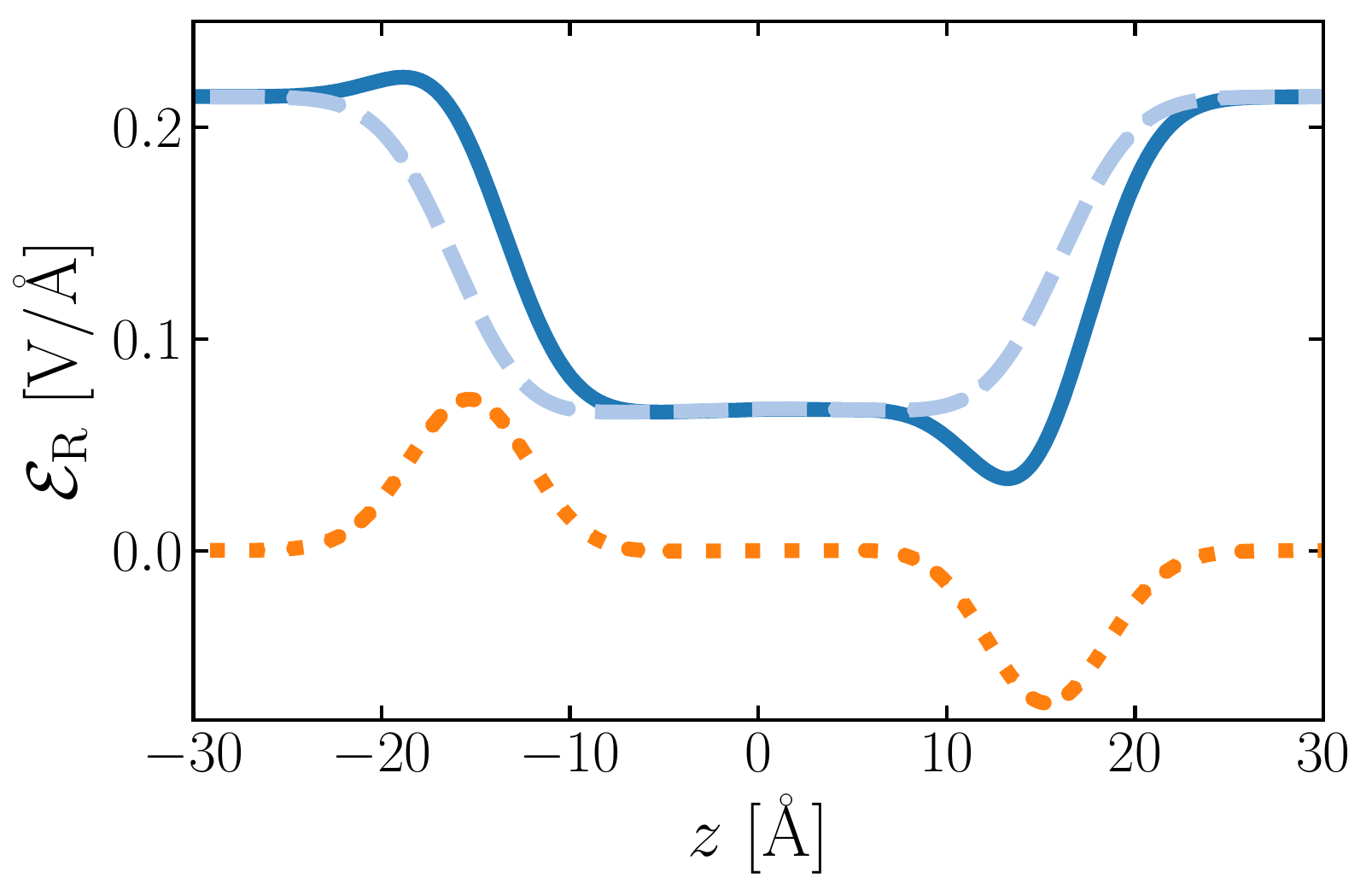}
  \caption{Accounting for the effects of spontaneous charge density on
    the restructuring field. The dotted orange line shows
    $\mathcal{E}_{\rm R}$ for the confined water system with
    $D=0$\,V/\AA, obtained from molecular simulation. The solid blue
    line shows $\mathcal{E}_{\rm R}$ with $D=0.15$\,V/\AA, also
    obtained from molecular simulation. The dashed line results from
    subtracting the dotted orange line from the solid blue line (this
    is also presented in \tcr{Fig.~4}).}
  \label{fig:SlabSuppl}
\end{figure*}

Even in the absence of an imposed electric or electric displacement
field, water exhibits a non-vanishing charge density close to the
interface. Consequently, the restructuring potential \tcr{Eq.~9} is
also non-vanishing. The corresponding restructuring field for the
confined system (see \tcr{Fig.~4}) at $D=0$\,V/\AA{} is shown by the
dotted line in Fig.~\ref{fig:SlabSuppl}. Also shown by the solid line
in Fig.~\ref{fig:SlabSuppl} is the full restructuring field at
$D=0.15$\,V/\AA. The DCT model presented in the main article aims to
describe the response of the confined system to uniform fields, and
does not account for effects of the spontaneous charge density at
$D=0$\,V/\AA{} on $\mathcal{E}_{\rm R}$. Thus, when comparing results
from molecular simulation to the DCT model, the restructuring field at
$D=0$\,V/\AA{} needs to be subtracted from the full restructuring
field for $D\neq 0$. This results in the dashed line presented in
Fig.~\ref{fig:SlabSuppl} and \tcr{Fig.~4}\,(c).

\section*{Predicting the variance of $M$ for confined water with LMFT}

The solid lines shown in \tcr{Fig.~4}\,(e) are not fits to the
simulation data. Rather, they show predictions of the simple DCT
model. The value \tcr{$\epsilon=71.7$} was obtained from the
simulations of bulk water (see \tcr{Fig.~2}). In order to find $w$,
the average variance $\overline{\langle M^2\rangle}_{\rm ew}$ from the
three Ewald simulations was calculated. Simple electrostatic arguments
then give
\begin{equation}
  w =
  \frac{4\pi\beta}{A}\left(\frac{\epsilon}{\epsilon-1}\right)\overline{\langle M^2\rangle}_{\rm ew},
\end{equation}
where $A$ is the cross-sectional area of the slab. For the system
under consideration in \tcr{Fig.~4}, this gives \tcr{$w=33.2$}\,\AA,
which was used in \tcr{Eqs.~10} \tcr{and~11} to produce
\tcr{Figs.~4}\,(c) and~(d). The Gaussian distributions shown by the
solid lines in \tcr{Fig.~4}\,(e) have zero mean, and variance
$\overline{\langle M^2\rangle}_{\rm ew}/\eta_L$.

\section*{Discussion of the Green's functions}

\tcr{Equation~4} gives the Green's function of Poisson's equation in
periodic space. This is well established
\cite{caillol1992asymptotic,de1986computer}, but it is useful to see
this in the present notation before going on to discuss SCA. To this
end, note that $\psi^{(\lambda)}$ is independent of $\kappa$. For
formal manipulations, it is convenient to consider the limit
$\kappa\to\infty$,
\begin{equation}
  \psi^{(\lambda)}(\mbf{r}) =
  \frac{1}{L^3}\sum_{\mbf{k}\neq 0}\frac{4\pi}{k^2}\me^{i\mbf{k}\cdot\mbf{r}} -
  \frac{2\pi\lambda}{3}\frac{|\mbf{r}|^2}{L^3}.
\end{equation}
Calculating the Laplacian gives
\begin{align}
  \nabla^2\psi^{(\lambda)}(\mbf{r}) &=
  -\frac{4\pi}{L^3}\sum_{\mbf{k}\neq 0} \me^{i\mbf{k}\cdot\mbf{r}} -
  \frac{4\pi\lambda}{L^3}, \\[7pt]
  &= -\frac{4\pi}{L^3}\sum_{\mbf{k}} \me^{i\mbf{k}\cdot\mbf{r}} -
  \frac{4\pi\lambda}{L^3} + \lim_{\mbf{k}\to\mbf{0}}\frac{4\pi}{L^3}\me^{i\mbf{k}\cdot\mbf{r}}, \\[7pt]
  &= -4\pi\left[\frac{1}{L^3}\sum_{\mbf{k}} \me^{i\mbf{k}\cdot\mbf{r}} + \frac{\lambda-1}{L^3} \right].
\end{align}
From the completeness relation,
$(1/L^3)\sum_{\mbf{k}}\exp(i\mbf{k}\cdot\mbf{r})$ is a set of
periodically replicated $\delta$-functions. It is then clear that
$\psi^{(\lambda)}$ is the electrostatic potential of a periodically
replicated unit point charge embedded in a uniform background charge
$(\lambda-1)/L^3$. Note that the uniform background charge has two
contributions: one from the $\lambda$-containing term in \tcr{Eq.~4},
and another from the regularization of the Ewald sum i.e. from
excluding the $\mbf{k}=\mbf{0}$ term in reciprocal space. Indeed, for
the familiar tin-foil boundary conditions ($\lambda=0$),
\begin{equation}
  \label{eqn:nabla-psi0-full}
  \nabla^2\psi^{(0)}(\mbf{r}) = -4\pi\left[\frac{1}{L^3}\sum_{\mbf{k}} \me^{i\mbf{k}\cdot\mbf{r}} - \frac{1}{L^3} \right].
\end{equation}
It will be shown shortly that it is this $1/L^3$ contribution to the
uniform background that gets distorted into a Gaussian charge
distribution by SCA, while the `$\lambda$ contributions' are
unaffected. It is the latter that give rise to terms in the
Hamiltonian proportional to $|\tbf{M}|^2$ (see
e.g. Ref.~\citenum{caillol1994comments}), such as when a constant
electric displacement field is imposed \cite{zhang2016computing1}.

Let us now consider a point charge $q$ at the origin, along with a
compensating Gaussian charge distribution, also centered at the
origin:
\begin{equation}
  n_{{\rm SCA},0}(\mbf{r}) = q\delta(\mbf{r}) - q\kappa\exp(-\kappa^2 r^2)/\sqrt{\pi}.
\end{equation}
The potential due to $n_{{\rm SCA},0}$ is
\begin{equation}
  \phi_0(\mbf{r}) = \frac{q\erfc(\kappa|\mbf{r}|)}{|\mbf{r}|}.
\end{equation}
Now consider the charge distribution to due a periodic array of
$n_{{\rm SCA},0}$,
\begin{equation}
  n_{\rm SCA}(\mbf{r}) = \sum_{\mbf{n}}n_{{\rm SCA},0}(\mbf{r}+\mbf{n}L).
\end{equation}
By linear superposition, this gives rise to a potential,
\begin{equation}
  \label{eqn:pot-per-unigauss}
  \phi(\mbf{r}) = \sum_{\mbf{n}} \frac{q\erfc(\kappa|\mbf{r}+\mbf{n}L|)}{|\mbf{r}+\mbf{n}L|}.
\end{equation}

Now let us return to the Green's function. In the main text,
\tcr{Eq.~5} was obtained by simply neglecting the sum in reciprocal
space in \tcr{Eq.~4}. Considering the arguments above leading to
Eq.~\ref{eqn:pot-per-unigauss} (setting $q=1$), and from the
uniqueness theorem of electrostatics,
\begin{equation}
  \nabla^2\psi_{\rm SCA}^{(\lambda)}(\mbf{r}) = -4\pi\left[n_{\rm SCA}(\mbf{r}) + \frac{\lambda}{L^3}\right].
\end{equation}
Thus $\psi^{(\lambda)}_{\rm SCA}$ is the Green's function for a
periodically replicated unit point charge with a compensating Gaussian
charge, \emph{and} embedded in a uniform charge distribution
$\lambda/L^3$. Again, it is instructive to consider tin-foil boundary
conditions explicitly,
\begin{equation}
  \label{eqn:nabla-psi0-lmf}
  \nabla^2\psi_{\rm SCA}^{(0)}(\mbf{r}) = -4\pi n_{\rm SCA}(\mbf{r}).
\end{equation}
Comparing Eqs.~\ref{eqn:nabla-psi0-full} and~\ref{eqn:nabla-psi0-lmf}
makes it apparent that under tin-foil boundary conditions, SCA alters
the charge distribution by distorting the compensating uniform
background charge (associated with every point charge in the system)
into a compensating Gaussian charge distribution.

\section*{Notes on the dielectric constant for a short-ranged system}

The purpose of this section is to present a perspective on how the
dielectric constant in a system with short-ranged electrostatic
interactions can be understood. For simplicity, an infinite system is
considered. To begin, consider the Maxwell equation for dielectrics,
relating the electric displacement $\mbf{D}$ to the Maxwell electric
field $\mbf{E}$ and polarization $\mbf{P}$:
\begin{subequations}
  \label{eqn:Maxwell}
  \begin{align}
    \mbf{D}(\mbf{r}) &= \mbf{E}(\mbf{r}) + 4\pi\mbf{P}(\mbf{r}),  \label{eqn:Maxwell-r} \\[5pt]
    \mbf{D}(\mbf{k}) &= \mbf{E}(\mbf{k}) + 4\pi\mbf{P}(\mbf{k}).  \label{eqn:Maxwell-k}
  \end{align}
\end{subequations}
For convenience, the relation has been given both in real space
(Eq.~\ref{eqn:Maxwell-r}) and reciprocal space
(Eq.~\ref{eqn:Maxwell-k}). The dielectric tensor $\bm{\epsilon}$
relates $\mbf{D}$ and $\mbf{E}$,
\begin{subequations}
  \label{eqn:constit}
  \begin{align}
    \mbf{D}(\mbf{r}) &= \int\!\mrm{d}\mbf{r}^\prime\,\bm{\epsilon}(\mbf{r},\mbf{r}^\prime)\cdot\mbf{E}(\mbf{r}^\prime), \label{eqn:constit-r} \\[5pt]
    \mbf{D}(\mbf{k}) &= \bm{\epsilon}(\mbf{k})\cdot\mbf{E}(\mbf{k}).  \label{eqn:constit-k}
  \end{align}
\end{subequations}
If one asserts that $\bm{\epsilon}$ is unchanged irrespective of
whether SCA or a full electrostatic treatment is used, then
Eq.~\ref{eqn:constit} suggests that one way to view any differences in
dielectric screening as due to differences in $\mbf{E}$ (and
consequently $\mbf{D}$). It is argued below that this viewpoint is
useful for understanding the properties of the LMFT/SCA system
presented in the main article.

It is useful to state some additional known results for the infinite
system in which a full electrostatic treatment is used (see
e.g. Ref.~\citenum{MaddenKivelson1984sjc}). Combining
Eqs.~\ref{eqn:Maxwell-k} and~\ref{eqn:constit-k} gives,
\begin{equation}
  \label{eqn:chi}
  4\pi\mbf{P}(\mbf{k}) = [\bm{\epsilon}(\mbf{k})-\mbf{1}]\cdot\mbf{E}(\mbf{k}) \equiv 4\pi\bm{\chi}(\mbf{k})\cdot\mbf{E}(\mbf{k}),
\end{equation}
where $\mbf{1}$ is the unit tensor. Both $\mbf{P}$ and $\mbf{E}$
depend upon the shape of the sample. In contrast, both $\bm{\epsilon}$
and $\bm{\chi}$ are shape-independent quantities. The polarization can
also be expressed in terms of the external field $\mbf{E}^{(0)}$,
\begin{equation}
  \label{eqn:chi0}
  4\pi\mbf{P}(\mbf{k}) = 4\pi\bm{\chi}^{(0)}(\mbf{k})\cdot\mbf{E}^{(0)}(\mbf{k}).
\end{equation}
The external susceptibility $\bm{\chi}^{(0)}$ is of interest as it is
directly related to molecular correlations in the system. As
$\mbf{E}^{(0)}$ is shape-independent, however, it immediately follows
that $\bm{\chi}^{(0)}$ depends upon sample shape. Relating
$\bm{\chi}^{(0)}$ to $\bm{\epsilon}$ is most readily achieved by
relating $\mbf{E}$ to $\mbf{E}^{(0)}$. For the infinite system
considered here,
\begin{equation}
  \label{eqn:E-E0-full}
  \mbf{E}(\mbf{k}) = \mbf{E}^{(0)}(\mbf{k}) - 4\pi\hat{\mbf{k}}\hat{\mbf{k}}\cdot\mbf{P}(\mbf{k}).
\end{equation}
Combining Eqs.~\ref{eqn:chi}, \ref{eqn:chi0} and~\ref{eqn:E-E0-full}
gives,
\begin{equation}
  \label{eqn:chi0-as-chi-full}
  \bm{\chi}^{(0)}(\mbf{k}) = [\mbf{1}+4\pi\hat{\mbf{k}}\hat{\mbf{k}}\cdot\bm{\chi}(\mbf{k})]^{-1}\cdot\bm{\chi}(\mbf{k}).
\end{equation}
For an isotropic system, if $\bm{\epsilon}$ is an intensive property
then it cannot depend upon the direction of
$\mbf{k}$. Thus,
\[\lim_{\mbf{k}\to\mbf{0}}\bm{\epsilon}(\mbf{k})
= \epsilon\mbf{1},\]
where $\epsilon$ is a scalar. This leads to the
familiar results for the transverse (perpendicular to $\hat{\mbf{k}}$)
and longitudinal (parallel to $\hat{\mbf{k}}$) components of
$\bm{\chi}^{(0)}$:
\begin{subequations}
  \label{eqn:chi-comp}
  \begin{align}
    \lim_{\mbf{k}\to\mbf{0}} 4\pi\chi^{(0)}_{xx}(\mbf{k})  =
    \lim_{\mbf{k}\to\mbf{0}} 4\pi\chi^{(0)}_{yy}(\mbf{k}) &= \epsilon-1,                   \label{eqn:chi0-xx} \\[5pt]
    \lim_{\mbf{k}\to\mbf{0}} 4\pi\chi^{(0)}_{zz}(\mbf{k}) &=  \frac{\epsilon-1}{\epsilon}. \label{eqn:chi0-zz}
  \end{align}
\end{subequations}
Following convention, a coordinate system has been chosen such that
$\hat{\mbf{k}}$ defines the $z$ direction.

For the SCA system, Eq.~\ref{eqn:E-E0-full} now reads
\begin{equation}
  \label{eqn:E-E0-sca}
  \mbf{E}(\mbf{k}) = \mbf{E}^{(0)}(\mbf{k}) - 4\pi[1-\exp(-k^2/4\kappa^2)]\hat{\mbf{k}}\hat{\mbf{k}}\cdot\mbf{P}(\mbf{k}).
\end{equation}
It is apparent that, at long wavelengths,
$\mbf{E}\approx\mbf{E}^{(0)}$, i.e. there is no depolarizing field in
the SCA system. Similarly, Eq.~\ref{eqn:chi0-as-chi-full} now reads
\begin{equation}
  \label{eqn:chi0-as-chi-sca}
  \bm{\chi}^{(0,{\rm SCA})}(\mbf{k}) =
        \left[\mbf{1}+4\pi[1-\exp(-k^2/4\kappa^2)]\hat{\mbf{k}}\hat{\mbf{k}}\cdot\bm{\chi}(\mbf{k})\right]^{-1}\cdot\bm{\chi}(\mbf{k}).
\end{equation}
This leads to,
\begin{equation}
  \label{eqn:chi0-sca}
  \lim_{\mbf{k}\to\mbf{0}} 4\pi\chi^{(0,{\rm SCA})}_{\alpha\alpha}(\mbf{k}) =  \epsilon-1, 
\end{equation}
with $\alpha = x, y$\,or $z$. Taking $\epsilon$ to be unchanged
between the full and SCA treatments for electrostatics,
Eq.~\ref{eqn:chi0-sca} appears in line with the results presented in
the main paper (see \tcr{Fig.~3} and Eq.~\tcr{15}).

\end{document}